\documentclass[fleqn,twoside,twocolumn,nofootinbib,showkeys]{revtex4} 
\usepackage[sec,nocpr]{ujp} 

\begin{document}
\title[Electrophysical Characteristics of Near-Surface Layers]
{ELECTROPHYSICAL CHARACTERISTICS\\ OF NEAR-SURFACE LAYERS IN
\boldmath$p$-Si
CRYSTALS\\ WITH SPUTTERED Al FILMS AND SUBJECTED\\ TO ELASTIC DEFORMATION}%
\author{B.V.~Pavlyk}
\affiliation{Ivan Franko National University of Lviv, Department of Electronics}
\address{107, Tarnawsky Str., Lviv 79017, Ukraine}
\email{pavlyk@electronics.wups.lviv.ua}
\author{M.O.~Kushlyk}%
\affiliation{Ivan Franko National University of Lviv, Department of Electronics}%
\address{107, Tarnawsky Str., Lviv 79017, Ukraine}%
\author{R.I.~Didyk}
\affiliation{Ivan Franko National University of Lviv, Department of Electronics}
\address{107, Tarnawsky Str., Lviv 79017, Ukraine}
\author{Y.A.~Shykorjak}%
\affiliation{Ivan Franko National University of Lviv, Department of Electronics}%
\address{107, Tarnawsky Str., Lviv 79017, Ukraine}%
\email{pavlyk@electronics.wups.lviv.ua}%
\author{D.P.~Slobodzyan}%
\affiliation{Ivan Franko National University of Lviv, Department of Electronics}%
\address{107, Tarnawsky Str., Lviv 79017, Ukraine}%
\email{pavlyk@electronics.wups.lviv.ua}%
\author{B.Y.~Kulyk\,}%
\affiliation{Ivan Franko National University of Lviv,\\
$~~$Scientific-Technical and
Educational Center of Low Temperature Studies}%
\address{50, Drahomanov Str., Lviv 79005, Ukraine}%

 \udk{???} \pacs{61.50.Ks, 61.72.Ff,\\[-3pt] 68.35.bg} \razd{\secvii}

\autorcol{B.V.\hspace*{0.7mm}Pavlyk, M.O.\hspace*{0.7mm}Kushlyk,
R.I.\hspace*{0.7mm}Didyk et al.}

\setcounter{page}{742}%

\begin{abstract}
The deposition of Al film onto the (111) surface of a \textit{p}-Si
crystal was shown to induce a deformation in the near-surface layer of
the latter. Provided that the crystal strain is elastic and
uniaxial, the gettering of defects in the near-surface layer is
observed, which is confirmed by a change in the dependence of
the specimen resistance on the elastic strain magnitude. The maximum
depth of the defect capture has been calculated on the basis of
the energy of interaction between the deformed layer and dislocations.
\end{abstract}
 \maketitle

\section{Introduction}

Our understanding of a deformation in silicon is mainly reduced to a
change in the equilibrium state of the silicon crystal lattice under
the action of external stresses. Owing to a modification of the
crystal lattice, the parameters of the electron band structure in
silicon also change. Variations in the mobility of charge carriers,
their capture cross-section, concentration of defects, and so on
result in a variation of  defects total electric conductivity, the
latter being strongly dependent on the strength, direction, and type
of a crystal \mbox{deformation.}

Mechanical stresses in a silicon substrate can also be induced when
a metallic or dielectric film is grown on its surface. Mechanical
stresses in the plane of contact between the film and the substrate
arise owing to a mismatch between the parameters of their lattices
\cite{1}. For instance, the mismatch between the crystal lattice
parameters for silicon and aluminum is about 25\%. Under the action
of a biaxial strain, the charge carriers become more mobile (as a rule,
by 50 to 70\%), which results in a reduction of the resistance in
the strained \mbox{crystal layer \cite{2}.}

During the formation of a film on the semiconductor surface, there
emerge regions of mechanical stresses, where defects of various
types can be localized. This region can also stimulate a generation
of dislocations, e.g., at the following plastic
de\-formation~\cite{3}.

This work aims at studying the influence of the deformation fields formed as a
result of the metal film sputtering onto the surface of \textit{p}-silicon
crystals subjected to uniaxial and elastic strain on the electro-physical
characteristics of those crystals.

\section{Experimental Technique}

Specimens to study were cut out from a wafer of single-crystalline silicon of
the \textit{p}-type (KDB-10 grade). After the standard procedures of cutting,
grinding, and chemical polishing, the both specimen facets (111) were covered
with Al film contacts 90~nm in thickness. The Al films were deposited in a
vacuum chamber VUP-5M at a pressure of 10$^{-2}$~Pa and a temperature of
350$~^{\circ}\mathrm{C}$ in such a way that the surface edges were covered by
the film, whereas its central part remained free \cite{4}. With the use of the
thermocompression welding technique, gold wires were attached to the Al films.
In their turn, the wires were soldered to the input contacts of a measuring unit.

Electroconductivity measurements were carried out on a vacuum
deformation installation at room temperature and a residual gas
pressure of 10$^{-2}$~Pa. The external strain was applied cyclically, in
the elastic deformation interval, along the crystallographic
direction [112] (i.e. in parallel to the lateral facets (111) and
(110) of a rectangular specimen) with the pressure forces up to
40~MPa and at deformation rates of 8 and 32~$\mu\mathrm{m/\min}$.
One cycle of uniaxial strain included the stages of specimen
squeezing, deformation relief, and specimen holding between those
stages at room temperature for a period varying from a few minutes
\mbox{to an hour.}

After the mechanically stimulated changes in the electric conductivity had
been measured, the aluminum film was etched. The obtained surface (111) of the
specimen was selectively etched, and then its structural researches were
carried out with the use of optical, electron, and atomic force
microscopies. Afterward, the studied surface was etched again, layer-by-layer
at a low etching rate, and examined on an optical microscope.

\section{Theoretical Calculations}

The crystal lattice of silicon has a diamond cubic structure with
the lattice parameter $a_{0}=$\linebreak $=0.5431$\textrm{~nm }and
the shortest interatomic distance equal to 0.24\textrm{~nm}. The
plane (111) is characterized by the closest packing of atoms, and
the corresponding calculated interatomic distance amounts to
0.375\textrm{~nm}. If the near-surface layer of single-crystalline
silicon contacts with a material, in which the lattice period
differs from that of silicon, there emerges a near-contact region,
the layers in which have intermediate values of lattice parameter.
As a result, such a mismatch induces mechanical stresses in the
single crystal described by a deformation potential. The symmetry of
crystal layers decreases at that, and two lattice parameters have to
be introduced. Below, the lattice constant in the (111) plane will
be denoted as $a_{\parallel}$, and that in the direction perpendicular to
the \mbox{plane (111) as $a_{\perp}$.}

Aluminum has a face-centered cubic packing, its crystal lattice parameter
amounts to 0.408\textrm{~nm}, and the shortest interatomic distance
equals 0.289\textrm{~nm}. Our computer simulation showed that atoms in the
aluminum lattice also have the closest packing characterized by the minimum
interatomic distance, as was for the (111) plane of silicon. Therefore, when
aluminum is deposited onto the silicon (111) surface, the aluminum atoms also
form the plane (111), because this arrangement is the most beneficial
energetically, since the deformations in the contacting crystal lattices are
minimal. It should be noted that, according to the results of work \cite{5},
the diffusion coefficient of Al atoms is close to zero at the indicated
temperatures of their sputtering onto the Si surface; therefore, the diffusion
of Al into Si was practically absent.

For silicon, which is a covalent crystal, the binding energy of
atoms in the crystal lattice is equal to the sum of the energies of
individual atoms, the energies of electrostatic interactions
(electron--electron, electron--nucleus, and nucleus--nucleus ones),
and the energy of exchange interaction emerging owing to the
electron exchange \cite{6,7}. Hence, the total energy \mbox{looks
like}
\begin{equation}
U(r)_{\rm Si}=E_{0}+\frac{K-A}{1-S^{2}},\label{eq1}%
\end{equation}
where $E_{0}$ is the energy of a single atom, $K$ the energy of electrostatic
interactions, $A$ the energy of exchange interaction, $S$ the integral of
non-orthogonality (its value falls within the interval from 0 to 1), and $r$
the interatomic distance. The corresponding binding energy for atoms in the
aluminum crystal lattice equals%
\begin{equation}
U(r)_{\rm Al}={\frac{1}{4\pi\varepsilon_{0}}}{\frac{\alpha e^{2}}{r}},\label{eq2}%
\end{equation}
where $\alpha$ is the Madelung constant, and $e$ the electron charge. Hence,
the magnitude of silicon atom displacement in the near-surface layer is
proportional to the force, with which the aluminum atoms act on it; it is also
true for aluminum atoms in the near-surface layer. According to the results of
work \cite{8}, the distance between Si and Al atoms amounts to
0.254\textrm{~nm}. Therefore, from formulas (1) and (2), we can calculate the
corresponding values for the displacements of atoms in the first atomic
layers. As a result, we obtain $r_{\mathrm{Si}}=-0.01$\textrm{~nm} and
$r_{\mathrm{Al}}=0.075$\textrm{~nm}.

\begin{figure}
\vskip1mm
\includegraphics[width=4cm]{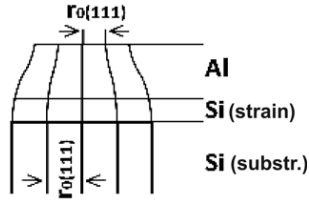}
\vskip-3mm\caption{Schematic diagram of a crystal lattice parameter
change in the heterostructure }
\end{figure}

Let the deformed region consist of three layers with different
thicknesses. These are (Fig. 1) the aluminum film with a deformed lattice
practically across the whole film thickness, a few atomic layers of silicon
with a deformed lattice, and the substrate with a non-deformed structure.
To make calculations simpler, let us assume that the strain is uniform across
the thickness of each layer, $h_{i}$; in other words, the averaged values of
atomic displacements from the corresponding equilibrium positions are used.
According to elasticity theory \cite{9}, the mechanical stress in a
uniform crystal film with the \textquotedblleft undeformed\textquotedblright%
\ lattice constant $a_{i}$ and thickness $h_{i}$ is determined by the
expression
\begin{equation}
\tau_{\bot i}=\frac{h_{\bot i}(E_{\bot i}(a_{\parallel i}-a_{\perp0}%
))}{(1-\nu_{\perp i})a_{\perp0}},%
\end{equation}
where $E_{i}$ and $\nu_{i}$ are the Young modulus and Poisson's ratio,
respectively, for the $i$-th layer. The relative deformation in each layer is
determined by a tensor, in which $\varepsilon_{12}=\varepsilon_{13}%
=\varepsilon_{23}=0$,
\begin{equation}
\varepsilon_{11}=\varepsilon_{22}=\frac{E_{2}h_{2}}{E_{1}h_{1}}
\frac{1-\nu_{2}}{1-\nu_{1}}\frac{a_{i+1}-a_{i}}{a_{i}a_{i+1}+a_{i}%
},\label{eq4}%
\end{equation}
and
\begin{equation}
\varepsilon_{33}=\frac{2\nu_{1}}{1-\nu_{1}}\frac{E_{2}h_{2}}{E_{1}h_{1}}%
\frac{1-\nu_{2}}{1-\nu_{1}}\frac{a_{i+1}-a_{i}}{a_{i}a_{i+1}+a_{i}%
}.\label{eq5}%
\end{equation}
Here, $h_{1}$ and $h_{2}$ are the average thicknesses, $a_{i}$ and
$a_{i+1}$ the average lattice parameters, $E_{1}$ and $E_{2}$
the Young moduli, and $\nu_{1}$ and $\nu_{2}$ Poisson's ratios for the
deformed substrate layer and the sputtered film, respectively.

The total energy of the heterostructure equals
\begin{equation}
W=\frac{1}{2}\int{\sum{\tau_{i}\varepsilon_{i}dV}}.%
\end{equation}
It consists of three components: the energy of mismatch deformations, the
energy of interaction between mismatch deformations and misfit dislocations,
and the energy of misfit dislocations. Provided that the initial conditions
for the formation of the epitaxial layer were so set that misfit dislocations are
not formed (in particular, a substrate temperature of 350$~^{\circ}\mathrm{C}$
and the absence of external stresses), the total energy is determined by the
expression
\begin{equation}
W=\frac{\tau Sh_{1}(2\varepsilon_{11}+\varepsilon_{33})}{2x}, \label{eq6}%
\end{equation}
where $S$ is the sputtered film area, and $x$ is the coordinate reckoned
into the crystal depth.

The energy of edge dislocations located in the crystal bulk before the epitaxial
sputtering is calculated by the formula%
\begin{equation}
W_{d}=\frac{NGb^{2}}{4\pi}\left\langle l\right\rangle \ln\frac{x}{r_{0}%
}, \label{eq7}%
\end{equation}
where $N$ is the concentration of dislocation outcrops on the surface of
silicon crystal, $G$ the shear modulus, $b$ the absolute value of Burgers
vector, $\left\langle l\right\rangle $ the average dislocation length, and
$r_{0}$ the dislocation core radius. If an external mechanical loading is
applied to the crystal, the energy of the near-surface contact layer changes
by the magnitude%
\begin{equation}
\Delta W=\pm\frac{F L}{S_{1} E}, \label{eq8}%
\end{equation}
where $F$ is the force applied to the crystal, $L$ the initial crystal length,
and $S_{1}$ the area of the crystal substrate, to which the force is applied. The
sign plus is selected if the external strain and the mechanical field under
the film simultaneously squeeze or stretch the crystal lattice. But if the
external loading acts oppositely to the near-surface strain field, the sign
minus should be selected.

\begin{figure}
\vskip1mm
\includegraphics[width=\column]{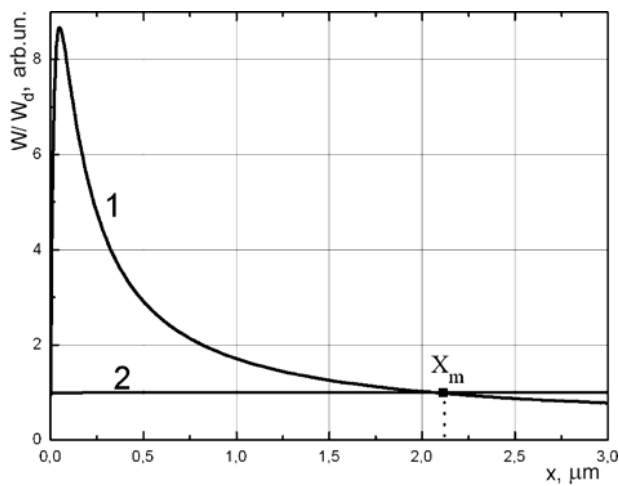}
\vskip-3mm\caption{Dependences of the total energy of deformed Al
and Si layers (\textit{1}) and the energy of dislocations
(\textit{2}) on the coordinate of a point in the crystal}
\end{figure}

The edge dislocation, in its turn, can bend and move in the direction
perpendicular to that external strain. The concentration of dislocation
outcrops on the surface ($N$, in $\mathrm{cm}^{-2}$ units) also varies at
that, and, consequently, the energy of dislocations in the near-surface region
changes proportionally to $N$.

The plot describing the dependence of the ratio between the energy of the deformed
layer and the energy of a dislocation in the course of external mechanical
loading on the coordinate is depicted in Fig.~2. One can see that the
corresponding curves intersect at the point $X_{m}$, the maximum depth of
defect capture by the near-surface layer from edge dislocations. Its specific
value depends on the magnitude of external strain and the parameters
describing the process of metal film sputtering. If the thickness of the deformed
layer increases, which corresponds to variations in the film sputtering
conditions (such as the substrate temperature, the sputtering rate, the duration
of annealing after the sputtering, and others), the position of point $X_{m}$
shifts into the crystal depth. All theoretical dependences were plotted for
defect-free crystals covered with an epitaxial Al film at the dislocation concentration up to
10--100~\textrm{cm}$^{-3}$.

\begin{figure}
\vskip1mm
\includegraphics[width=\column]{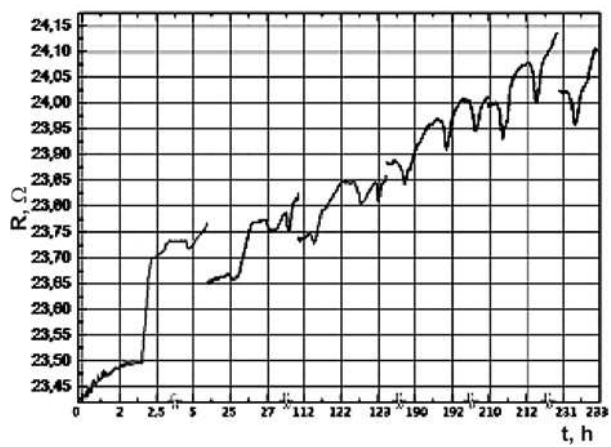}
\vskip-3mm\caption{Time dependences of the resistance of specimen~1
in the squeezing--loading relief cycles. Between the cycles, the
specimen was held at room temperature for 18.75 (\textit{1}), 91.2
(\textit{2}), 65.25 ({\it 3}), 18.5 (\textit{4}), and 17.4~h
(\textit{5}) }\vskip3mm
\end{figure}

\begin{figure}
\includegraphics[width=\column]{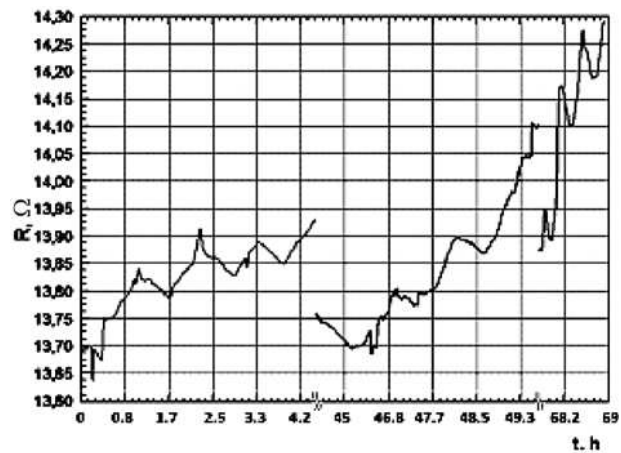}
\vskip-3mm\caption{The same as in Fig.~1, but for specimen~2.
Between the cycles, the specimen was held at room temperature for 41
(\textit{1}) and 17.5~h (\textit{2})  }
\end{figure}

\begin{figure}
\vskip1mm
\includegraphics[width=7.5cm]{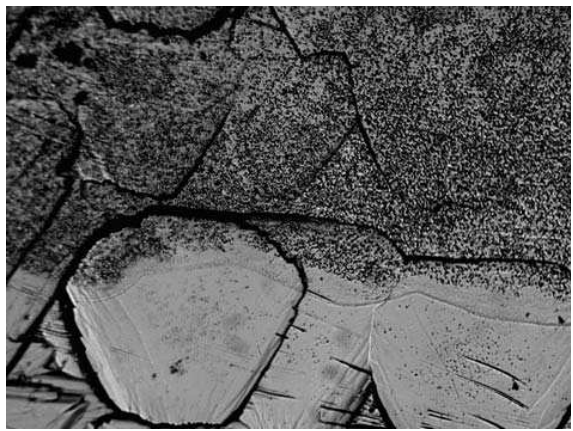}
\vskip-3mm\caption{Microphoto of the surface (111) of silicon
covered with an Al film made with the help of an optical microscope
at 70-fold magnification }\label{fig.5}\vskip3mm
\end{figure}

\begin{figure}
\includegraphics[width=\column]{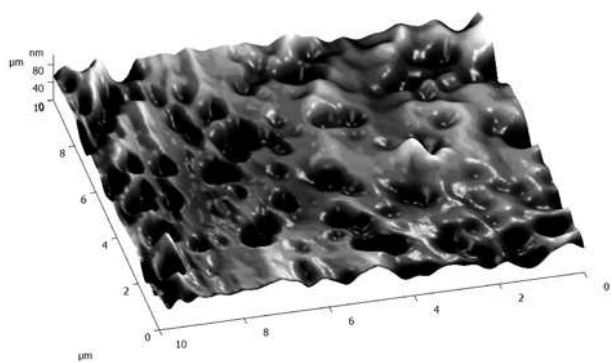}
\vskip-3mm\caption{Microphoto of the surface (111) of silicon
covered with an Al film made with the help of an atomic force
microscope }\label{fig.6}\vskip3mm
\end{figure}

\begin{figure}[h!]
\includegraphics[width=\column]{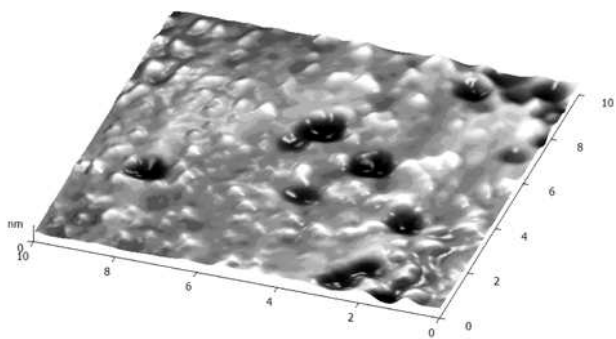}
\vskip-3mm\caption{The same as in Fig.~6 but for silicon not covered
with an Al film  }
\end{figure}

\section{Experimental Results}

Experimentally, we studied two specimens: a specimen, which was elastically
deformed beforehand by applying a squeezing force of 55 MPa (specimen~2) and a
specimen, which was not preliminarily subjected to a deformation (specimen~2).
In Figs. 3 and 4, the variations in the dependences of the specimen
electroresistances on the elastic strain time and the specimen
annealing duration at room temperature are shown for both specimens. The results of
experiments in the course of a single \textquotedblleft squeezing--strain
relief\textquotedblright\ cycle are presented by a solid curve; the break between
the curves indicates the holding of the specimen after the previous
\textquotedblleft squeezing--strain relief\textquotedblright\ cycle (for the break
duration, see the figure caption). The cycle following the holding corresponds
to the following plot interval (a solid curve), the initial section of which
(the resistance growth) corresponds, as a rule, to the squeezing stage, and
the following recession stage corresponds to the strain relief. These
dependences make it evident that (i)~the magnitude of residual specimen
electroresistance grows from cycle to cycle, with the intensity of this growth
decreasing after several cycles; (ii)~in the specimen preliminarily subjected
to a deformation, the processes that are responsible for the gradual increase in
the specimen electroresistance are slowed down; (iii)~after the mechanical
loading relief and the holding of the specimen at room temperature for not more than
an hour, the specimen demonstrates a slow growth of its electroresistance; and
(iv)~holding the specimen at room temperature for 24~h and longer gives rise
to a partial restoration of its conductance.

The structural researches consisted in a microscopic examination of
the specimen surface state after the selective etching. Figure~5
shows the photos obtained on an optical microscope. They demonstrate
that an increased concentration of defects is formed under the
sputtered film after the processes of squeezing and strain relief.
The analysis of the photos (Fig.~6) of the selectively etched
surface obtained with the use of atomic force microscopy testifies
that those defects are groups of structural surface defects that
differ from dislocation etching pits by their depth and edge shape.
The authors of work \cite{10} regarded those defects as the clusters
of point defects.

At the same time, such an increase in the defect concentration was not
observed on the substrates of crystals with sputtered Al films but not
subjected to the action of any external force (Fig.~7). This fact allows us to
assert that those defects were gettered in the near-surface layer of the
crystal from the crystal bulk, as a result of the external strain and the
presence of the Al film.

Since dislocations create a field of mechanical stresses around themselves, they are
efficient drains for any defects or impurities in the crystal. When an
external mechanical field is applied to the specimen, dislocations can wander
through the lattice (in particular, loop-like dislocations with fixed ends are
capable of bending), so that an additional capture of defects from the crystal
bulk takes place. In the case of uniaxial crystal strain, dislocations
approach the near-surface layer with mechanical stresses induced by a mismatch
between the lattice parameters of silicon and aluminum film, and defects
localized around the dislocation core become captured in the disturbed
near-surface layer. This model explains a variation of the specimen conductance
during a deformation. If the loading decreases, the dislocation gets straight
and returns back to the initial position, whereas some captured defects remain
in the deformation potential field induced by the lattice parameter mismatch.
As a result, the specimen resistance changes between the deformation cycles.

The defect wandering through the lattice can induce the appearance of traps that
can capture charge carriers. The filling of those traps is accompanied by a
reduction in the concentration of charge carriers and, respectively, in a
growth of the specimen resistance after the deformation has been terminated. The
role of such defects can be played by dislocations that release the captured
defects to the near-surface layer, by vacancies taken out by dislocations to
the surface, and so forth.

However, the explanation of the variation in the specimen
electroconductance is not limited to a single mechanism. The blocking of
donor centers that have arisen or shifted in the course of
deformation is also possible. One more mechanism consists in a
variation of the magnitude of charge carrier scattering by defects
taken out to and captured in the \mbox{near-surface layer.}

After carrying out the researches dealing with the level-by-level etching of
the surface, we obtained a layer with the concentration of etching pits equal
to the initial value (before the sputtering of the Al film). In such a way, we
experimentally determined the maximum depth of probable defect capture by the
near-surface deformed layer, which turned out to equal 1.5~$\mu\mathrm{m}$.
This value coincides with the theoretically calculated one to within the
calculation error.

\section{Conclusions}

The sputtering of a metal film onto the silicon substrate and the subsequent
elastic strain of the latter result in the defect capture from the crystal
bulk in the near-surface deformed layer. Our calculations showed that the
near-surface layer of silicon is deformed owing to the mismatch between the
lattice parameters in the crystal and the film. As a result, there emerges a
deformation potential near the surface, which promotes the defect capture (the
Cottrell atmosphere) from dislocations that approach the surface under the
action of an external elastic strain. The analysis of the microscopic structure
of the surface showed that those defects are the clusters of point defects.
Possible mechanisms of the influence of the elastic strain on the crystal
electroconductivity by means of increasing the defect concentration in its
near-surface region are described. Theoretical calculations for the energy
distribution across the deformed near-surface layer and the energy of
dislocations are carried out, and the maximum depth of probable defect
capture is found, which correlates well with the experimentally obtained results.

\rezume{%
Б.В. Павлик, М.О. Кушлик, Р.І. Дідик,\\ Й.А. Шикоряк, Д.П.
Слободзян, Б.Я. Кулик}{ЕЛЕКТРОФІЗИЧНІ ХАРАКТЕРИСТИКИ\\
ПРИПОВЕРХНЕВИХ ШАРІВ КРИСТАЛІВ Si\\ $p$-ТИПУ, З НАПИЛЕНИМИ ПЛІВКАМИ
Al, \\ПІДДАНИХ ПРУЖНІЙ ДЕФОРМАЦІЇ} {Показано, що осаджена плівка Al
на поверхню (111) кристала Si($p$) формує деформаційне поле в
приповерхневому шарі. За одновісної пружної деформації кристала
спостерігається гетерування дефектів з об'єму зразка у
приповерхневому шарі під напиленою плівкою. Отримана залежність
зміни величини опору цих зразків від величини пружної деформації
підтверджує гетерування електрично активних дефектів у
приповерхневому деформованому шарі. Проведено теоретичні розрахунки
максимальної глибини захоплення цих дефектів на основі енергії
взаємодії деформованого шару та дислокацій.}

\end{document}